# Tunable Ampere phase plate for low dose imaging of biomolecular complexes

**Amir H. Tavabi[1], Marco Beleggia[2], Vadim Migunov[1], Alexey Savenko[3], Sara Sandin[4], Ozan Öktem[5], Rafal E. Dunin-Borkowski[1], Giulio Pozzi[1, 6]**

1- Ernst Ruska-Centre for Microscopy and Spectroscopy with Electrons and Peter Grünberg Institute, Forschungszentrum Jülich, D-52428 Jülich, Germany
2- Center for Electron Nanoscopy, Technical University of Denmark, 2800 Kgs Lyngby, Denmark
3- FEI Company, Achtseweg Noord 5, 5600 KA, Eindhoven, The Netherlands
4- School of Biological Sciences, Nanyang Technological University, 60 Nanyang Drive, Singapore 637551, Singapore
5- Centre for Industrial and Applied Mathematics, Department of Mathematics, KTH - Royal Institute of Technology, Stockholm, Sweden
6- Department of Physics and Astronomy, University of Bologna, Viale B. Pichat 6/2, 40127 Bologna, Italy

*A novel device that can be used as a tunable support-free phase plate for transmission electron microscopy of weakly scattering specimens is described. The device relies on the generation of a controlled phase shift by the magnetic field of a segment of current-carrying wire that is oriented parallel or antiparallel to the electron beam. The validity of the concept is established using both experimental electron holographic measurements and a theoretical model based on Ampere's law. Computer simulations are used to illustrate the resulting contrast enhancement for studies of biological cells and macromolecules.*

The imaging of biomolecular complexes with near-atomic spatial resolution is crucial for our understanding of how biological functions emerge from a set of building blocks that are carefully orchestrated by the laws of physics and chemistry. Aberration correction in transmission electron microscopy (TEM) now provides sub-Ångström point resolution at low and medium (20-300 kV) accelerating voltages [1-3]. When accompanied by the latest generation of electron detectors, these instruments provide an almost ideal platform for addressing outstanding problems in structural biology. However, one issue remains: biological structures interact with an incoming high-energy electron beam by modulating the phase of the electron wave very weakly. A "phase-sensitive" imaging method is then required to turn an invisible phase modulation into visible contrast that can be recorded, analyzed and linked back to the number, positions and species of atoms present.

In-focus phase contrast devices [4, 5] that are analogues of the optical phase plate (PP) introduced by Zernike [6] have been investigated for decades and emerged from a wide range of possible phase-sensitive imaging techniques as promising and viable methods for enhancing the contrast of biological specimens in the TEM. However, according to Glaeser [5], most PPs that have been proposed so far suffer from deficiencies that include insufficient contrast enhancement, additional aberrations, short lifetime, lack of a straightforward alignment method and/ or electrostatic charging by the electron beam (even when self-charging is responsible for producing phase contrast [7 & 8]). In particular, charging is very difficult to measure and control, deteriorates device performance and limits widespread applications.

Here, we introduce a new substrate-free PP concept for TEM that is based on Ampere's law and addresses all of these problems. We refer to the device as a "tunable Ampere phase plate" (TAPP). It is designed to provide almost-ideal phase contrast, while maintaining 1-4 Å spatial resolution and providing both tunability and ease of application.

The functionality of the TAPP is provided by a magnetic field circulating around a vertical segment of current-carrying wire, which adds a position-dependent phase shift to a passing electron wave. When the TAPP is positioned in the back focal plane of the imaging lens, it acts as an additional transfer function that enhances phase contrast from the object.

We used focused ion beam (FIB) milling to fabricate a prototype TAPP from etched Au wires in the form of three orthogonal segments, thereby making a hook-shaped device, as shown in Fig. 1a. One segment of hook could then be positioned parallel and the other two perpendicular to the incident electron beam direction. In the present study, the device was mounted in the specimen plane in order to measure its phase shift. The geometry of the wire is shown in Figs 1b and 1c, viewed at tilt angles of 70° and 0° with respect to the incident electron beam direction, respectively. The wire was connected electrically inside the TEM using a nanopositioning specimen holder, closing the circuit and allowing a current to flow. The phase shift introduced by the TAPP was then recorded (with the short segment of the hook oriented parallel to the electron beam direction) for varying currents using off-axis electron holography [9] and compared with a theoretical model based on Ampere's law.

Figures 1d - 1f show maps of the phase shift introduced by the TAPP, measured using electron holography and displayed in the form of 8x amplified phase contours. The tilted bar in each image is the shadow of the horizontal segments of the hook, which is opaque to the electrons, while the wider region at its center corresponds to the position of the vertical segment of the wire. Figure 1d shows a phase image recorded using electron holography in the absence of current flow. The signal is essentially flat, with only Fresnel fringes caused by the edges of the biprism wire that was used to form the hologram visible at the edge of the field of view, indicating that neither magnetic nor electric fields are present around the device. Although this conclusion is at first sight expected as there is no current flowing, its experimental verification is important as it demonstrates that there is no unwanted electron beam induced charging effect when the phase plate is operated inside the microscope. Together with its tunability, this is one the most advantageous aspects of the TAPP when compared with most other phase plate concepts. Figure 1e shows a phase image recorded when a nominal current of 2 mA is passed through the device. The rings represent projected magnetic field lines. The lack of perfect circular symmetry is attributed to the use of a perturbed reference wave in electron holography in the presence of a long-range magnetic field [10]. However, this imaging artifact does not influence the operation of the TAPP. Colors are used to represent the direction of the projected magnetic field. In accordance with Ampere's law, increasing the current two-fold to 4 mA (Fig. 1f) yields twice the phase shift. A comparison between Figs 1e and 1f reveals that at every point the magnitude of the signal is doubled, while maintaining an identical decay with distance (see Supplementary Information).

Figures 1g - 1i show corresponding calculated phase images, also displayed as 8x amplified contour maps, for a vertical current segment that has a length of 2 µm and for current flows of 0, 2 and 4 mA in Figs 1g, 1h and 1i, respectively. The black bar mimics the opaque projection of the device. The consistency between the simulated and experimental images confirms that the TAPP functions as an electron phase shifting device that is easily tunable.

In order to assess the performance of the TAPP device in the back focal plane of the microscope, we carried out simulations for a nucleosome core particle (NCP), PDB-ID: 1AOI, whose molecular structure is shown in Fig. 2a [11]. In order to mimic a realistic experimental setup, the NCP was virtually embedded in a 50-nm-thick layer of water ice and imaged at an accelerating voltage of 200 kV. Figures 2b and 2c show the amplitude and phase of the electron wave transmitted through the NCP, respectively, calculated using the approach described in Ref. [12]. The calculated amplitude image, which varies locally by only 2.5%, reveals that the NCP does not contribute substantially to the effective absorption of the ice embedment, which removes as much as 40% of the incoming electrons. In contrast, the phase image is more directly related to the local projected atomic/molecular weight of the NCP and reveals a relatively strong phase shift between the ice (1.59 rad, corresponding to a mean inner potential of 4.4 V for the 50-nm-thick ice layer at 200 kV) and the interior of the NCP (1.71 rad maximum), reflecting the presence of elements that are heavier than oxygen, as well as denser regions. Comparing Fig. 2a with Figs 2b and 2c visually, a resemblance is recognized more easily with the phase image, indicating that, as expected, structural information is encoded more directly in the phase of the electron wave than in its amplitude.

In comparison, Figs 2d - 2f show simulated Fresnel defocus image calculated in the absence of noise and aberrations, illustrating how the conventional out-of-focus technique blurs information when a defocus value that is large enough to provide sufficient contrast is used. At a defocus $Z$ of 0.05 µm, the contrast in Fig. 2d is very low and the image resembles more the amplitude in Fig. 2b than the phase in Fig. 2c. When the defocus is set to $Z = 0.5$ µm (Fig. 2e), the contrast increases to approximately 10% and some phase information is visible, but the structural details are now blurred. An increase in the defocus to $Z = 1.0$ µm (Fig. 2f) provides no further advantages, as the contrast decreases and the details are washed out.

Figures 2g - 2i show image simulations calculated for the TAPP device, revealing a clear and almost-ideal phase contrast image with high contrast of approximately 12% when a current of 0.60 mA is applied (Fig. 2h). Such an image cannot be achieved using the out-of-focus technique. At lower current values, for example 0.25 mA (Fig. 2g), phase contrast is preserved but the contrast is lower. An increase in the current to 3 mA results in a contrast reversal of the NCP, which now appears brighter than the background (Fig. 2i), as well as mixing phase and amplitude information and adding a surrounding halo. The "tail" of the NCP (marked by a red arrow in Figs. 2a and 2c) remains well defined in all of the TAPP simulations, indicating that 2-3 Å spatial resolution can be preserved even at relatively large current values. In contrast, when using the out-of-focus technique, the tail disappears as soon as the defocus value is high enough to provide sufficient contrast.

In summary, we have described a new substrate-free phase plate, which is based on Ampere's law and offers significant advantages over previous designs. It provides tunability, phase shifting homogeneity, minimal aberrations, almost no delocalization, ease of operation, almost ideal phase contrast for a weak phase object and a simplified fabrication procedure. In contrast to other proposals for phase plates based on magnetic fields, it does not rely on the presence of ferromagnetic material and can therefore be used in the strong objective lens field of the electron microscope. The implementation of a TAPP in a modern cryo electron microscope promises to lead to major breakthroughs in structural biology, ultimately in combination with spherical and chromatic aberration correction and electron tomography, to achieve 1-2 Å spatial resolution and the full three-dimensional atomic structure of complex biomolecules.

**Acknowledgments**

The research leading to these results has received funding from the European Research Council under the European Union's Seventh Framework Programme (FP7/2007-2013)/ ERC grant agreement number 320832.

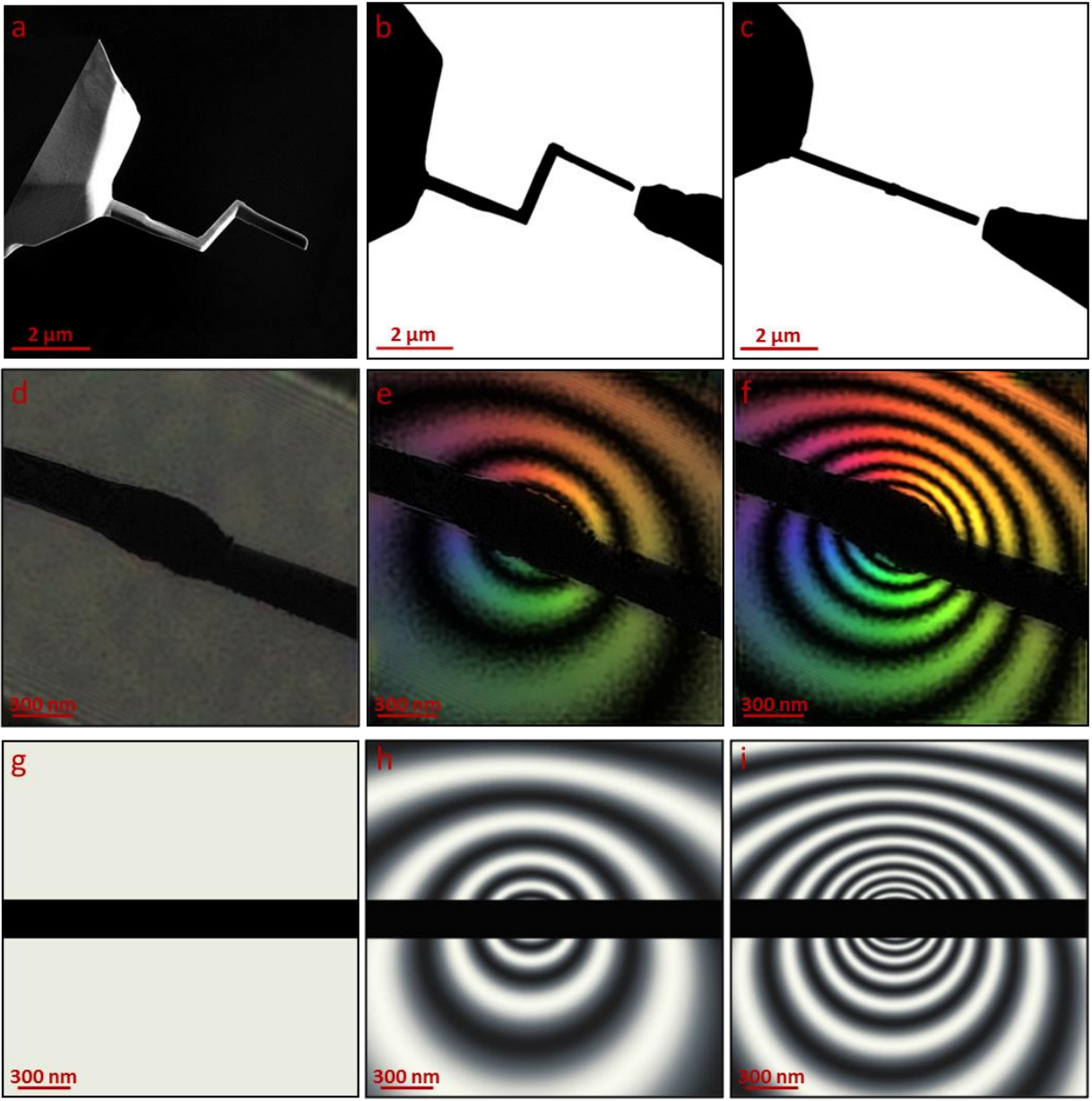

**Figure 1** a) Secondary electron image of the hook-shaped TAPP recorded before connecting it to a counter-electrode. b) and c) Bright-field images of the device viewed in the specimen plane in a TEM at tilt angles of 70° and 0° with respect to the optic axis of the microscope, respectively. The counter-electrode, which is moveable in the present setup, is visible on the right of each image. d)-f) 8x amplified phase images of the vacuum region around the phase plate recorded using off-axis electron holography for currents through the device of d) 0, e) 2 and f) 4 mA. g) - i) Simulated 8x amplified phase images for currents of g) 0, h) 2 and i) 4 mA, including the influence of the perturbed reference wave.

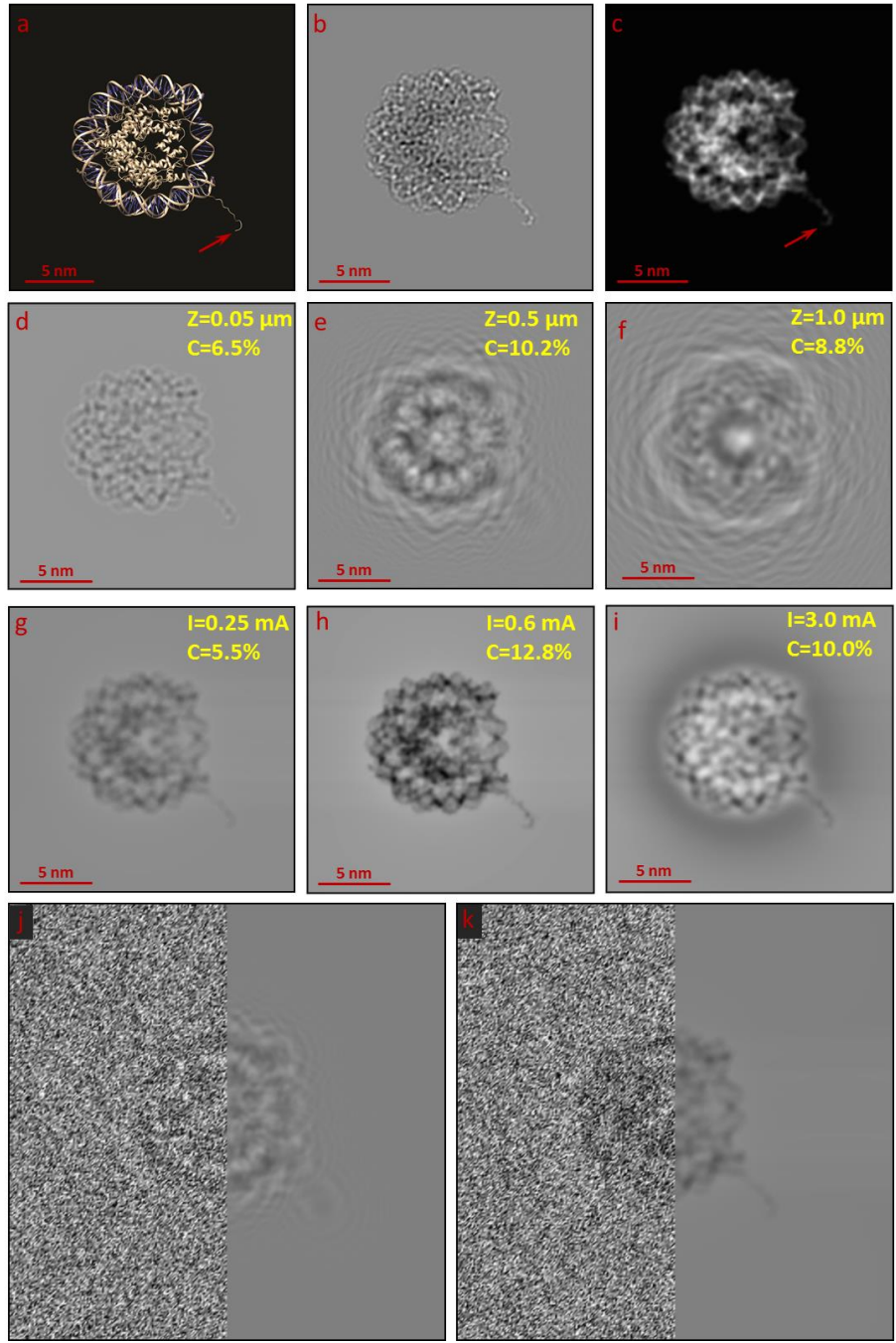

**Figure 2** a) Molecular structure of the NCP. b) Amplitude and c) phase of the electron wave transmitted by the NCP at 200 kV; the images are density-plotted over their full ranges: 0.770-0.790 for the amplitude and 1.59-1.71 for the phase. d)-f) Simulated out-of-focus Fresnel images for defocus values of d) 0.05, e) 0.5 and 1.0 µm. g)-i) Simulated TAPP images for currents of g) 0.2, h) 0.6 and i) 3.0 mA. The image contrast C, which is defined as $(I_{max}-I_{min})/(I_{max}+I_{min})$, is specified at the top right corner of each simulation. j) and k) show how typical levels of noise would affect the object visibility for the out-of-focus and TAPP images.